\documentstyle[12pt]{article}

\newcommand{\om}{\omega}
\newcommand{\al}{\alpha}

\newcommand{\cO}{{\cal O}}

\newcommand{\deebar}{\bar{\partial}}

\newcommand{\df}{\stackrel{\rm def}{=}}

\newcommand{\msc}[1]{\mbox{\scriptsize #1}}
\newcommand{\dsp}{\displaystyle}

\newcommand{\br}{\mbox{{\bf R}}}
\newcommand{\bz}{\mbox{{\bf Z}}}

\newcommand{\cA}{{\cal A}}

\newcommand{\ket}[1]{{|#1\rangle}}

\newcommand {\eqn}[1]{(\ref{#1})}

\newcommand{\cleqn}{\setcounter{equation}{0}}
\makeatletter
\@addtoreset{equation}{section}

\makeatother

\begin{document}
\vskip 7mm
%%% Title page %%%%%
\begin{titlepage}
 
 \renewcommand{\thefootnote}{\fnsymbol{footnote}}
 \font\csc=cmcsc10 scaled\magstep1
 {\baselineskip=14pt
 \rightline{
 \vbox{\hbox{hep-th/9905004}
       \hbox{UT-833}
       }}}

 \vfill
 \baselineskip=20pt

\bigskip

\begin{center}
\noindent{\Large \bf Multi-Strings on $AdS_3\times S^3$ 
                     from Matrix String Theory} 
\bigskip
\bigskip

\noindent{\it \large Kazuo Hosomichi and Yuji Sugawara} \\
{\sf hosomiti@hep-th.phys.s.u-tokyo.ac.jp~,~
sugawara@hep-th.phys.s.u-tokyo.ac.jp}
\bigskip

{\it Department of Physics, Faculty of Science, \\
\medskip
          University of Tokyo,\\
\medskip
     Bunkyo-ku, Hongo 7-3-1, Tokyo 113-0033, Japan}
\bigskip
\bigskip

\end{center}
 \vfill
 \vskip 0.5 truecm

%%%%%%%%%%%%%%%%%%%%%%%%%%%%%%%%%%%%%%%%%%%%%%%%%%%%
\begin{abstract}
   We analyze the Coulomb branch of Matrix string theory in the
 presence of NS5-branes. If we regard the components of $U(1)$ gauge
 fields as the dualized longitudinal coordinates, we obtain the
 symmetric product of $AdS_3\times S^3\times {\bf R}^4$ as the geometry
 of Coulomb branch. We observe that the absence or presence of the
 nonzero electric flux determines whether the string propagates
 in bulk as an ordinary closed string or is forced to live near the
 boundary. 

   We further discuss the issues of the physical spectrum from the
 viewpoint of Matrix string theory. We show that the twisted sectors
 of CFT on the symmetric orbifold, which correspond to glued strings,
 turn out to yield many chiral primaries that were hitherto considered
 to be missing. We also comment on the threshold energy in Liouville
 sector where continuous spectrum begins.
\end{abstract}

\setcounter{footnote}{0}
\renewcommand{\thefootnote}{\arabic{footnote}}
\end{titlepage}
\newpage

\section{Introduction}
\cleqn 

   The theory of gravity or string theory on three-dimensional anti
 de-Sitter space ($AdS_3$) recently has been a topic of considerable
 interest. It has been understood for a long time that the gravitational 
 degrees of freedom in the bulk of $AdS_3$ may be described by a
 conformal field theory on the boundary. After Maldacena presented 
 the celebrated conjecture on the duality between the gravity on the
 $AdS$ space and CFT on its boundary\cite{9711200}, the case of $AdS_3$
 achieved particular interest, as we know CFT in two dimensions best.
 It may be possible to prove the Maldacena's conjecture in more detail
 in the case of $AdS_3/CFT_2$. Much work has been done along this
 path\cite{9806194}--\cite{9904024}.

   In string theory, the $AdS_3$ space arises as the near-horizon
 geometry of the bound states of D1/D5-branes. We take its
 S-dual and consider the string theory on an NS1/NS5-brane background,
 in order to avoid the difficulty in considering string theories on RR
 backgrounds. The duality between string theory on $AdS_3$ and a certain
 boundary CFT has been discussed extensively. The generators of boundary
 superconformal algebras were directly constructed out of vertex
 operators on the superstring worldsheet\cite{9806194}. There have been
 attempts to study the spectrum of chiral primaries and observe
 the correspondence to some extent\cite{9812027,9812100}.
 
   There have been further attempts to realize the boundary CFT Ward
 identities in which the $AdS_3$ string theory was treated in
 a different way\cite{9812046}. Although there was some confusion
 between these two approaches, some recent works\cite{9903219,9903224}
 put these concepts in order. Now it turns out that the string theory
 on $AdS_3$ has two distinct sectors corresponding to two different
 configurations of the worldsheet. They are called as 
 the ``short string'' sector and the ``long string'' sector. The long
 strings are forced to live near the boundary with some fixed winding
 number, while the short strings can have arbitrary configurations of
 worldsheet and are free to propagate in bulk. The two sectors lead to
 two different explanations for the origin of the central charge of the
 boundary CFT, as was discussed in \cite{9903224}. The main purpose of
 the present paper is to propose a unified framework including both the
 short and long string sectors, based on Matrix string
 theory\cite{9703030}.

   One of the unresolved problems regarding the spectrum of chiral
 primaries is the mismatch of its upper bound. In the string theory on
 $AdS$ space it is of order $\sim k$, where $k$ is the number of
 $5$-branes. However, it is argued that the bound becomes of
 order $\sim pk$ in the boundary CFT, where $p$ is the number of
 $1$-branes. As was stated in our previous paper\cite{9812100}, it is
 natural to think that this discrepancy is due to the fact that we are
 only considering the theory of a single string on $AdS_3$. We expect
 that the multi-string system based on Matrix string theory may solve
 this inconsistency. In this paper we show that this is indeed the case.  

~

   This paper is organized as follows. We begin with the analysis of
 the Coulomb branch of Matrix string theory in the presence of
 NS5-branes. We are then lead to the sigma model on the moduli space of
 the Coulomb branch, which is identified as the second quantized string
 theory on  $AdS_3 \times S^3 \times \br^4$. Twisted sectors of this
 orbifold CFT correspond to glued strings, which are strings made
 from some fundamental strings glued together.\footnote{
   Hereafter we shall use the term ``{\em glued string}'' for strings
   which are made by gluing some fundamental strings together. We will
   keep the term ``{\em long string}'' for different objects, which
   appear in \cite{9903219,9903224}.}
 The long/short string sectors are distinguished by the presence/absence
 of nonzero electric flux on the worldsheet.  
 
   We shall also discuss the spectrum of chiral primaries from the
 viewpoint of Matrix string theory, especially the issue of missing
 chiral primaries and the threshold for continuous spectrum pointed
 out in \cite{9903224}. 

   Throughout this paper  we denote NS$5$-brane charge as $k$ and
 NS$1$-brane charge as $p$, following the convention of \cite{9806194}.

~

\section{Analysis of Gauge Theory}
\cleqn

   Matrix string theory of type IIA superstring\cite{9703030} is defined
 as the large $N$ limit of ${\cal N}=(8,8)$ $U(N)$ SYM theory in two
 dimensions. To incorporate $k$ longitudinal NS5-branes extending along
 the $016789$-directions, we add $k$ hypermultiplets belonging to the
 fundamental representation of $U(N)$. In ${\cal N}=(4,4)$ language
 the system consists of a $U(N)$ vector multiplet and hypermultiplets
 belonging to one adjoint and $k$ fundamental representations of $U(N)$.
 In the Coulomb branch the gauge group
 is generically broken down to $U(1)^N$. The massless scalar fields
 parametrizing the moduli space are $N$ abelian vectormultiplets, which
 correspond to the string coordinates along the directions transverse
 to the NS$5$-branes ($2345$ directions), and the $N$ neutral hypermultiplets
 corresponding to the coordinates along $6789$ directions. Usually the
 $6789$ directions are compactified on $T^4$ or $K3$.
 But in this article we would like to ignore
 the subtleties concerning the compactification of Matrix string theory
 and focus mainly on the six-dimensional part.
 
   One can obtain the exact metric of moduli space by one-loop analysis.
 This was done in \cite{9703031,9707158} for the case of $N=1$, and 
 similar calculation works also in the case of generic $N$. 

   In ${\cal N}=(2,2)$ terminology, an abelian vector multiplet consists
 of a chiral and a twisted-chiral multiplets, $(\Phi,\Sigma)$.
 We can write down the most generic action for these multiplets as an
 integral of a function $K(\Phi,\bar{\Phi},\Sigma,\bar{\Sigma})$ over
 superspace\cite{gates}. The condition for $K$ to give a ${\cal N}=(4,4)$
 supersymmetric gauge theory of vectormultiplets is given by
\begin{equation}
  K_{\Phi^i\bar{\Phi}^j}+K_{\Sigma^i\bar{\Sigma}^j}=0~,~~~~
  K_{\Phi^i\bar{\Phi}^j}-K_{\Phi^j\bar{\Phi}^i}=0
\end{equation}
   The metric and $B$ field on the target space are given by
\begin{eqnarray}
  ds^2&=& K_{\Phi^i\bar{\Phi}^j}d\Phi^i d\bar{\Phi}^j
         -K_{\Sigma^i\bar{\Sigma}^j}d\Sigma^i d\bar{\Sigma}^j \\
  B   &=&-\frac{1}{4}(
          K_{\Phi^i\Sigma^j}d\Phi^i d\Sigma^j
         -K_{\Phi^i\bar{\Sigma}^j}d\Phi^i d\bar{\Sigma}^j
         -K_{\bar{\Phi}^i\Sigma^j}d\bar{\Phi}^id\Sigma^j
         +K_{\bar{\Phi}^i\bar{\Sigma}^j}d\bar{\Phi}^id\bar{\Sigma}^j)
\end{eqnarray}
   Taking $Spin(4)$ and permutation symmetry into account, 
 we can conclude that the most generic form for $K$ is given by
\begin{eqnarray}
  K   &=& \sum_i K_i  \nonumber \\
  K_i &=& a (\Phi^i\bar{\Phi}^i-\Sigma^i\bar{\Sigma}^i)
         +b \left(
            \ln\Phi^i\ln\bar{\Phi}^i
           -\int^{\frac{\Sigma^i\bar{\Sigma^i}}{\Phi^i\bar{\Phi}^i}}
            \frac{d\xi}{\xi}\ln(1+\xi)
            \right)
\end{eqnarray}
   The terms with coefficients $a$ and $b$ are tree and one-loop
 contributions, respectively, and we can fix them as
 $a=\frac{1}{2g^2}$ and $b=\frac{k}{4\pi}$.

   We would like to note here that the one-loop contributions to the
 effective action arise only from the fundamental hypermultiplets, so
 that the resultant contribution is proportional to $k$. The one-loop
 contribution of adjoint fields cancel each other. This is to be
 expected, because the system has a larger supersymmetry (with $16$
 supercharges) when the fundamental hypermultiplets are absent.

   The effective action has the following bosonic part
\begin{eqnarray}
   S  &=& \sum_i S_i  \nonumber \\
   S_i&=& -\int d^2x
     \left(\frac{1}{g^2}+\frac{k}{2\pi y_{(i)}^2}\right)
     \left[\frac{1}{2}\partial_\mu y_{(i)}^p\partial^\mu y_{(i)}^p
     +\frac{1}{4}f_{(i)\mu\nu}f_{(i)}^{\mu\nu}\right] \nonumber \\&& 
     +\int_{\cal B} \frac{k}{6\pi y_{(i)}^4}
      \epsilon_{pqrs}dy_{(i)}^p dy_{(i)}^q dy_{(i)}^r y_{(i)}^s 
\end{eqnarray}
\[
 i=1,\ldots,N~,~~~
 \mu,\nu=0,1~,~~~
 p,q,r,s=2,3,4,5
\]
 Here $y_{(i)}^p$ and $f_{(i)\mu\nu}$ are the four scalars and the
 $U(1)$ field strength of the $i$-th abelian vectormultiplet.
 ${\cal B}$ is some open three-dimensional space whose boundary is 
 the worldsheet of Matrix string. In Maldacena's near horizon limit or
 the weak coupling limit of IIA superstring theory, we send 
 $g^{-2} \equiv g_s^2 l_s^2$ to zero and drop the tree term from the
 action. Then the metric of the target space is no longer asymptotically
 flat. The four-dimensional spatial directions now have a tube metric,
 and we expect to obtain the $AdS_3\times S^3$ geometry by
 incorporating the longitudinal directions. After changing the coordinates
 from $y_{(i)}^p$ to the radial coordinates $y_{(i)}$ and the $SU(2)$
 group elements, the action becomes
\begin{equation}
  S_i=-\frac{k}{2\pi}\int_{\cal M}d^2x\left[
  \frac{1}{2y_{(i)}^2}\partial_\mu y_{(i)}\partial^\mu y_{(i)}
 +\frac{1}{4y_{(i)}^2}f_{(i)\mu\nu}f_{(i)}^{\mu\nu}
  \right]
 + (SU(2)\mbox{~WZW with level~}k )_i
\end{equation}
 The full supersymmetric extension of this action is then
\begin{equation}
   S_i=-\frac{k}{4\pi}\int_{\cal M}d^2xd^2\theta Y_{(i)}^{-2}DY_{(i)}DY_{(i)}
 + (SU(2)\mbox{~SWZW with level~} k )_i
\label{eff-action}
\end{equation}
 Here $Y_{(i)}$ is an ${\cal N}=(1,1)$ superfield which has $y_{(i)}$
 as the lowest component. Its $\theta$-expansion reads
\[
 Y_{(i)}=y_{(i)}+i\theta\psi_{(i)}
 +\frac{i}{4}\theta\theta\epsilon^{\mu\nu}f_{(i)\mu\nu}.
\]
 Note that the level of the bosonic part of SWZW in
 \eqn{eff-action} should be shifted to $k-2$ by means of a chiral
 rotation so as to make the fermionic fields free, as was discussed in
 \cite{9707158,Allen,CHS}. Hence the level of the total current,
 including the fermionic contribution, should remain $k$.

~

\section{Short Strings and Long Strings}
\cleqn

   Various interesting phenomena occur when we compactify the spatial
 direction on $S^1$. Although the gauge field in two dimensions has no
 dynamics, it carries some topological informations, according to which
 we can classify the sectors of Matrix string theory.
 
   Various sectors can be labeled by the periodicity conditions along
 $S^1$. As was explained in the original papers on Matrix string
 theory\cite{9703030}, we can twist the periodicity of fields which take
 values in the Cartan subgroup of $U(N)$ by elements of the Weyl group,
 namely, the $N$-th permutation group. The twisted sectors are
 pictorially understood as glued strings, or some strings mutually glued
 together into a single string. For example, there is a sector which is
 labeled by the $N$-th cyclic permutation. This sector is interpreted
 as the sector where $N$ fundamental strings are joined together to form
 a glued string. As we shall explain later, the glued strings
 constructed in this way are accountable for the missing chiral primaries.

   Another interesting phenomenon is the presence of nonzero $U(1)$
 electric flux. In the flat background with no NS5-brane charge, this flux
 describes the existence of D0-branes\cite{9703030}, and in the
 T-dualized framework it describes the $(p,q)$-string
 sectors\cite{9510135,9705029}. However, as our discussion below is
 based on the effective action \eqn{eff-action}, the presence of
 electric flux may have a different interpretation. We shall show below
 that the string becomes ``long'' if there is nonzero electric flux on its
 worldsheet.

   To simplify the problem, we focus on the case of a single string 
 (the case $N=1$) for the time being. To see the implication of nonzero
 electric flux, let us write down the relevant terms in the effective
 action below;
\begin{equation}
  S= -\frac{k}{4\pi}\int_{\Sigma}\, d^2x \, 
  \left(\partial_\mu\phi\partial^\mu\phi +\frac{1}{2}e^{-2\phi}
  f^{\mu\nu}f_{\mu\nu}\right)+\ldots
  \label{action1}
\end{equation}
 Here the ``Liouville field'' $\phi$ is defined as $y=e^{\phi}$ and 
 the worldsheet $\Sigma$ is a cylinder or the two-punctured sphere.
 We can quantize the system canonically in the $A_0=0$ gauge.
 The only dynamical variable is the Wilson line
 $\dsp U=\exp \left(\oint A_1 \,dx\right)$. $\psi(U)\sim U^n$ is
 the eigenfunction of the electric field strength
 $\dsp \Pi \equiv e^{-2\phi}E$, which is the canonical conjugate of the
 Wilson line. Note that the eigenvalue is quantized due to
 the periodicity of the Wilson line. When there is nonzero electric
 flux, it yields the following contribution to the effective action
 (here we move to the Euclidean signature):
\begin{equation}
  S \sim \int_{\Sigma} \, d^2x \, e^{2\phi_0}n^2
  ~,~~~~ n\in {\bf Z}
  \label{action2}
\end{equation}
 To minimize the action, we must have
 $\phi_0\sim -\infty~ (\mbox{or~} y=0)$
 if there is nonzero electric flux. On the contrary, $\phi_0$ is
 arbitrary if there is no electric flux. Hence we come to the conclusion 
 that the string is forced to live near the source NS$5$-branes if there 
 is nonzero electric flux on the worldsheet.

 In the present situation, the NS$5$-branes are located at the origin,
 $y=0$. If we T-dualize the system, the parametrization of the
 radial direction is reversed, namely, $y=0$ and $y=\infty$ are
 interchanged.

   We would like to see the correspondence between the effective action
 of Matrix string theory and the action of superstring on
 $AdS_3\times S^3$ in some detail. To do this, we shall first move to
 the Euclidean worldsheet. As was obtained in the previous section, the
 effective field theory has the following ``Liouville part'',
\begin{equation}
  S=\frac{1}{8\pi}\int_{\Sigma}d^2x\left(
  \partial\varphi\bar{\partial}\varphi+e^{-\sqrt{\frac{2}{k}}\varphi}
  f_{\mu\nu}f_{\mu\nu}
  \right),
\label{action 3}
\end{equation}
 where $\varphi=\sqrt{2k}\phi$ up to a constant shift. To ensure the
 conformal invariance, we must add the background charge term,
\[
  \frac{Q}{8\pi}\int_{\Sigma}\varphi R_{(2)},
\]
 to the above action. The background charge $Q$ must be determined such
 that the term $\dsp e^{-\sqrt{\frac{2}{k}}\varphi}f_{\mu\nu}f_{\mu\nu}$
 has the correct conformal dimension. Then the effective action is given by
\begin{eqnarray}
  S&=~&~~(\mbox{super Liouville theory
               with the background charge~}Q)\nonumber \\
   &~ &+\,(SU(2)\mbox{~SWZW with level~}k)\nonumber \\
   &~ &+\,(\mbox{SCFT on~}{\bf R}^4 )
\label{action 4}
\end{eqnarray}
 This theory has the central charge
\begin{equation}
  c_{\rm total}=\left(1+3Q^2+\frac{1}{2}\right)
               +\left(\frac{3(k-2)}{k}+\frac{3}{2}\right)
               +6
\end{equation}
 Below we show that one must choose $Q$ differently for long and short
 strings. The observation of \cite{9903224} that the CFT on the short
 and long strings describe respectively the Coulomb branch and the Higgs
 branch near the small instanton singularity.

~

\noindent
\underline{{\em 1. Short String~} ($n=0$)}  

   In this sector it is appropriate to assign the canonical dimension
 $0$ to the gauge field $A_\mu$. The background charge is fixed by this 
 condition as $Q=-\sqrt{\frac{2}{k}}$. The central charge becomes
 $c=12$.

   It is easy to see the correspondence of this CFT and the world-sheet
 theory of short string on $AdS_3$ background. The worldsheet action
 of a fundamental string on $AdS_3$ is given by \cite{9806194};
\begin{equation}
  S=\frac{1}{8\pi}\int_{\Sigma}d^2x\left(
   \partial\varphi\bar{\partial}\varphi
  +e^{\sqrt{\frac{2}{k}}\varphi}\deebar \gamma \partial \bar{\gamma}\right) ,
\label{AdS3 sigma model}
\end{equation} 
 or equivalently, by introducing the auxiliary fields $\beta$ and
 $\bar{\beta}$ we have;
\begin{equation}
  S=\frac{1}{8\pi}\int_{\Sigma}d^2x\left(
   \partial\varphi\bar{\partial}\varphi
  -e^{-\sqrt{\frac{2}{k}}\varphi}\beta\bar{\beta}
   +\beta \deebar\gamma +\bar{\beta}\partial\bar{\gamma}\right) .
\label{AdS3 sigma model 2}
\end{equation}
 The short string theory on $AdS_3$ is defined on an arbitrary compact
 Riemann surface, because it should describe arbitrary propagation and
 interaction of closed strings in bulk. The $(\beta , \gamma)$-system
 should have the conformal dimensions $(1,0)$ as in Wakimoto
 representation\cite{Wakimoto}. From these facts we see the operator
 $\oint \gamma^{-1}\partial\gamma$ cannot take nonzero classical value.
 The RNS superstring action on $AdS_3$ background is obtained as the
 supersymmetric extension of the above action. In quantizing the system,
 we can choose the light-cone gauge $\gamma \sim z$, $\psi^+=0$
 ($z$ is a holomorphic coordinate on $\Sigma$ and
  $\psi^+$ is the fermionic coordinate along the longitudinal direction)
 of \cite{9812216,9903224} and eliminate the longitudinal degrees of
 freedom. After this gauge fixing, we find that the CFT of the
 transversal degrees of freedom coincides with the one we have obtained
 from the gauge theory analysis, including the value of the background
 charge of the Liouville part.

   Here we make an important remark. The gauge condition $\gamma\sim z$
 is quite different from the nonzero classical value of
 $\dsp \oint\gamma^{-1}\partial\gamma$. One must not confuse these two
 distinct notions. If we impose the light-cone gauge condition on a
 string theory defined on a generic Riemann surface, we should impose it
 on each of the {\it local} coordinate patches. Even in flat background the
 light-cone gauge cannot be imposed globally unless we take a
 cylindrical worldsheet. 

~

\noindent
\underline{{\em 2. Long String~} $(n\neq 0)$}  

 As we have seen above, in the case of $n\neq 0$, the worldsheet of the
 string must be near the NS$5$-branes, $y\sim 0$. The nonzero
 electric flux $\dsp \Pi\equiv e^{-\sqrt{\frac{2}{k}}\varphi} E$ also
 indicates that the gauge field $A_z$ and $\bar{A}_{\bar{z}}$
 should have the ``geometric dimension'' $(1,0)$ and $(0,1)$, and the
 operator $\dsp e^{-\sqrt{\frac{2}{k}}\varphi}$ should have the conformal
 weight $(-1,-1)$. This leads to 
 the Liouville background charge $\dsp Q= \sqrt{2k}-\sqrt{\frac{2}{k}}$. 
 The central charge now becomes $c=6k$.

   This sector is identical to the CFT on the ``long string'' in the
 light-cone gauge, which is discussed in \cite{9812216,9903224}. As was
 discussed in these papers, we can bosonize the eight free fermions in
 this theory to define eight spin fields, which are fermions in the sense 
 of the worldsheet as well. Using these spin fields we can construct a
 ${\cal N}=(4,4)$ superconformal algebra with $c=6k$. It is the sum of
 the two superconformal algebras arising from the ${\bf R}^4$ part and
 the remaining part, with $c=6$ and $c=6(k-1)$, respectively.

   Now we would like to compare our CFT with the CFT on the long string
 worldsheet. Recall that Matrix string theory describes the IIA strings
 and the (short or long) strings on $AdS_3 \times S^3$ are the objects
 in IIB string theory. So as to make a comparison we have to apply
 T-duality to the two-dimensional field theory.

   Let us start from IIB side. The relevant part of the world-sheet action
 is given by \eqn{AdS3 sigma model}, which we rewrite with
 $\gamma = \gamma^0+i\gamma^1$. Of course $\gamma^0$, $\gamma^1$
 are the time and spatial coordinates parameterizing the boundary of
 Euclidean $AdS_3$. We can partially gauge-fix the conformal symmetry
 by the condition $\gamma^0=t$, and obtain
\begin{equation}
  S=\frac{1}{8\pi}\int_{\Sigma}d^2x\left\{
   \partial\varphi\bar{\partial}\varphi
  +e^{\sqrt{\frac{2}{k}}\varphi}\left((\partial_t \gamma^1)^2
+ (\partial_x \gamma^1)^2 + 2\partial_x \gamma^1 \right)  \right\} ,
\label{AdS3 sigma model 3}
\end{equation}
 We can T-dualize this action according to the standard procedure(see,
 for example, \cite{T-duality}), with respect to the $U(1)$ isometry 
 $\gamma^1 ~ \longrightarrow ~\gamma^1 + \alpha$
\begin{equation}
  S=\frac{1}{8\pi}\int_{\Sigma}d^2x\left\{
   \partial\varphi\bar{\partial}\varphi
  +e^{-\sqrt{\frac{2}{k}}\varphi}\left((\partial_t \tilde{\gamma}^1)^2
+ (\partial_x \tilde{\gamma}^1)^2 \right) 
+ 2i \partial_t \tilde{\gamma}^1   \right\} ,
\label{AdS3 sigma model 4}
\end{equation}
 where $\tilde{\gamma}^1$ denotes the dual coordinate of $\gamma^1$.

   On the other hand, our effective action of Matrix string
 theory\eqn{action 3} can be rewritten, if a suitable source term
 ensuring nonzero electric flux on the worldsheet is added, as follows;
\begin{eqnarray}
  S &=& \frac{1}{8\pi}\int_{\Sigma}d^2x\left\{
  \partial\varphi\bar{\partial\varphi}+e^{-\sqrt{\frac{2}{k}}\varphi}
  \left((\partial_tA_1)^2+(\partial_xA_1)^2\right)\right\}
  +2i\left(\oint_{t=\infty}-\oint_{t=-\infty} A_1dx\right) 
    \nonumber \\
    &=& \frac{1}{8\pi}\int_{\Sigma}d^2x\left\{
  \partial\varphi\bar{\partial\varphi}+e^{-\sqrt{\frac{2}{k}}\varphi}
  \left((\partial_tA_1)^2+(\partial_xA_1)^2\right)+2i\partial_tA_1
  \right\}
\label{AdS3 sigma model 5}
\end{eqnarray}
 where we have chosen the $A_0=0$ gauge, and added a gauge-fixing term
 with respect to the residual gauge symmetry. Action
 \eqn{AdS3 sigma model 5} is clearly the same as \eqn{AdS3 sigma model 4}
 if we make the identification $\tilde{\gamma}^1= A_1$.

   Some comments are in order. First of all, long strings have nonzero
 winding number $\dsp \langle\oint\gamma^{-1}d \gamma \rangle =1$, which 
 implies that $\gamma$ should be of conformal dimension $-1$. This
 assignment of conformal weight forces us to impose the condition
 $\gamma\sim z$ {\it globally}. This is possible only when the
 worldsheet is cylindrical. Hence long strings must have cylindrical
 worldsheets, as was mentioned previously. Note also that
 this assignment of conformal weight is consistent with the interpretation
 that $\gamma$ and $A_\mu$ are dual to each other. 

   Moreover, the term
 $\dsp \sim 2e^{\sqrt{\frac{2}{k}}\varphi}\partial_x\gamma$ in
 \eqn{AdS3 sigma model 3} plays the role of source term for the
 winding number of $\gamma$, while the corresponding term
 $\dsp \sim 2i \partial_t A_1$ in the dual action \eqn{AdS3 sigma model 5}
 is no other than the source term for the electric field
 $\Pi \equiv e^{-\sqrt{\frac{2}{k}}\varphi} E$. Therefore we can say
 that the nonzero winding number of $\gamma$ in IIB theory side precisely
 corresponds to nonzero electric flux in the IIA Matrix string theory side. 

   In the above argument we can see explicitly, as stated previously, the
 interchange of $y=0$ and $y=\infty$ under the T-duality transformation.
 Hence a long string, which is forced to live near $y\sim 0$ in IIA
 picture, is at $y\sim +\infty$ in IIB picture.    

   The argument given above was based on a proposal of \cite{9707093}
 that we should assign the canonical dimension $0$ to the gauge field
 $A_\mu$ in the Coulomb branch CFT, while we should assign the geometrical
 dimension $1$ in the Higgs branch. Although originally we have been
 analyzing the Coulomb branch, the nonzero electric flux requires the
 redefinition of the conformal dimension of the gauge field, which then
 leads us to the physics of the Higgs branch. At the same time the
 object becomes a long string, whose dynamics suitably describes the
 Higgs branch CFT near the small instanton singularity, as was discussed 
 in \cite{9903224}.
 
\section{Twisted Sectors}
\cleqn
~
   Matrix string theory contains various sectors in which some fundamental 
 strings are glued together. The field contents on the worldsheet of
 glued strings are the same as those of a single string, so we can
 determine whether a glued string is long or short in the same way as
 with a single string. 
   A generic sector of Matrix string theory corresponds to a set of glued
 strings, each of which has an arbitrary length. Some of them are long,
 i.e. there is nonzero electric flux on their worldsheet. The sum of the 
 length of the long strings is the NS$1$-brane charge $p$, which is fixed.
 By the argument of the charge conservation, two strings can be glued
 together only when there is equal electric flux on the two
 worldsheets\cite{9705029}. In particular, a long string and a short string
 cannot be glued together. Then we are led to the following conformal
 field theory for this system:
 $\dsp Sym^{N-p}({\cal M}_{\msc{short}})\times Sym^p({\cal M}_{\msc{long}})$,
 where ${\cal M}_{\msc{short}} ({\cal M}_{\msc{long}})$ stands for the
 theory of a single short(long) string, as described in the previous
 section. Here we mean by $Sym^{N}({\cal M})$ the $N$-th symmetric
 product of a conformal field theory ${\cal M}$. This operation is
 defined in a similar manner as the sigma model on a symmetric orbifold
 ${\cal M}^N/S_N$ is defined from the sigma model on ${\cal M}$.

   Matrix string theory contains various twisted sectors describing many
 glued strings. So we expect that all of the connected and
 disconnected worldsheets having arbitrary genera can be reproduced in
 the large $N$-limit. Moreover, we further expect that the integration over
 the moduli of the worldsheets corresponds precisely to the large $N$
 limit of Matrix string theory. This was conjectured in \cite{9703030},
 and more detailed analyses were given in \cite{BBN} based on the
 interpretation of the BPS instantons in Matrix string theory as several
 plane curves. Although the analyses of \cite{BBN} were done in the flat
 background, this remarkable claim may be true even in the presence of
 NS$5$-branes. In this way we might be able to recover the full second
 quantized IIB superstring theory on $AdS_3\times S^3 \times {\bf R}^4$
 for the short string sector in the large $N$-limit. 

   How about the long string sector? Since the NS$1$-brane charge $p$ is
 fixed,  we have at most $p$-glued long string sector. This means that
 we cannot consider a worldsheet of a long string with arbitrarily large
 genus (although we assume $p>>1$). Furthermore the moduli of the
 worldsheet should be frozen, because the worldsheet of a long string
 must be cylindrical, lying near the boundary of $AdS_3$ in the IIB
 picture. Therefore it is more appropriate to regard the CFT of the
 long string sector as the boundary CFT itself, rather than the theory
 of gravity or string theory in bulk $AdS_3$.

   In our previous paper\cite{9812100} we treated the first quantized
 RNS superstring on $AdS_3\times S^3 \times T^4$ with the covariant
 formalism. We constructed a BRST invariant state that corresponds to
 the vacuum of the spacetime SCFT for the case of $p=1$. As for $p>1$
 no such state was found.  That observation fits nicely with the results
 obtained in this paper.

   We can see that if there were a {\em single} string carrying a
 winding charge greater than $1$, it would lead to an inconsistency. 
 In fact, the assumption
 $\dsp \langle \oint \, \gamma^{-1}d\gamma \rangle =p$ or $\gamma \sim z^p$
 implies that $(\beta,\gamma)$ have conformal weights $(p+1,-p)$,
 hence we have to improve the stress tensor as follows: 
\begin{equation}
  T_{\rm total}=T_{\rm worldsheet} + p\partial J^3.
\end{equation}
 This stress tensor would give rise to the central charge $c= 6kp^2$, if
 the worldsheet theory were originally a critical string theory. Hence
 this naive argument cannot explain the correct central charge unless
 $|p|=1$.

   In addition, our previous arguments have shown that the shift in the
 conformal weight of the gauge field, $A$, due to the electric flux is
 always $1$, {\em no matter how large the flux is.} In the IIB picture
 the conformal weight of $\gamma$ is always $-1$ when there is a nonzero
 electric flux in IIA picture, regardless of how large it is. So the
 winding number of a single long string is always $1$.

   So, the winding charge $p>1$ must be carried by some glued long
 strings. They must be composed of precisely $p$ single long strings,
 which leads to the CFT of $Sym^p({\cal M}_{\msc{long}})$ with the
 correct central charge $c=6kp$. We believe that the Matrix string
 theory is the most natural framework to explain why such an orbifold
 CFT arises. 

~

\section{Spectrum in the Long String Sector}
\cleqn

   In this section we discuss the spectrum of chiral primaries in the
 long string sector. As was given in the previous section, the CFT of
 the long string sector is given by $Sym^p({\cal M}_{\msc{long}})$,
 where ${\cal M}_{\msc{long}}$ is the CFT of a single long string.
 ${\cal M}_{\msc{long}}$ is the product of three CFT's, namely, a super
 Liouville theory with $\dsp Q=\sqrt{2k}-\sqrt{\frac{2}{k}}$, an SWZW
 theory with level $k$ and the superconformal sigma model on
 ${ M}^4$.\footnote{
    Previously we assumed that $M^4={\bf R}^4$. In this
 section we shall somehow generalize the situation and consider $M^4$
 as one of ${\bf R}^4, T^4$ or $K3$. Of course, as we have mentioned
 previously, if we consider the
 generalization for the choice of the internal manifold at the stage of
 Matrix string theory, there must arise the subtleties concerning the
 compactification of Matrix theory on higher-dimensional torus. But in
 this section we shall forget these difficulties and simply replace
 ${\bf R}^4$ by $T^4$ or $K3$.}
 The CFT on ${ M}^4$ and the remaining part have ${\cal N}=(4,4)$
 superconformal algebras with the central charge $c=6$ and $c=6(k-1)$,
 respectively. According to
 this product structure, any chiral primary of ${\cal M}_{\msc{long}}$
 can be written as the product of a chiral primary of ${ M}^4$ part and
 the chiral primary of the remaining part. We denote them as follows:
\begin{equation}
{\cal O}(\om, l) = {\cal O}_{\om}\,{\cal O}_l , ~~~ 
(\forall \om \in H^*({ M}^4), ~ l = 0, \ldots, k-2).
\label{spectrum 1}
\end{equation}
 Here ${\cal O}_{\om}$ is one of the chiral primaries of the ${ M}^4$
 part. They are labeled by the cohomology element, $\omega$, of ${ M}^4$.
 ${\cal O}_l$ is a chiral primary of the remaining part, which has the
 following form\cite{9903224,sl2};
\begin{equation}
  {\cal O}_l\df e^{l\sqrt{\frac{2}{k}}\varphi}V_l~,~~~
  (l=0,\ldots,k-2).
\label{vertex}
\end{equation}
 where $V_l$ is the highest weight operator in bosonic $SU(2)_{k-2}$ WZW
 theory with spin $l$, characterized by the following OPE with the
 bosonic $SU(2)$ currents
\begin{equation}
\left\{
\begin{array}{lll}
  k^3(z)V_l(0)& \sim &\dsp \frac{l}{2z}\,V_l(0) \\
  k^+(z)V_l(0)& \sim & 0
\end{array}
\right.
\end{equation}
 The chiral primary $\cO(\om,l)$ has the following quantum numbers:
\begin{equation}
  h=j= \frac{q(\om)+l}{2}, ~~~ \bar{h}=\bar{j}= \frac{\bar{q}(\om)+l}{2},~~~
  (\om \in H^{q(\om)\, \bar{q}(\om)} (M^4)).
\label{spectrum 2}
\end{equation}
 Alternatively, we can consider the corresponding Ramond vacuum
 $\ket{\om,l}$ that is obtained by spectral flow:
\begin{equation}
\begin{array}{ll}
\dsp  L_0 \ket{\om,l} = \frac{k}{4} \ket{\om,l} & \dsp
 K^3_0 \ket{\om,l} = \left(\frac{q(\om)}{2}+\frac{l}{2}-\frac{k}{2}\right)
     \ket{\om,l} , \\
\dsp  \bar{L}_0 \ket{\om,l} = \frac{k}{4} \ket{\om,l} & \dsp 
 \bar{K}^3_0 \ket{\om,l} = \left(\frac{\bar{q}(\om)}{2}
      +\frac{l}{2}-\frac{k}{2}\right)
     \ket{\om,l}.
\end{array}
\label{spectrum 3}
\end{equation}

  Now let us turn to the analysis for whole long string sector,
 $Sym^p({\cal M}_{\msc{long}})$. First of all, the spectrum for the
 untwisted sector is essentially the same as that for the single long
 string ${\cal M}_{\msc{long}}$.
 We again exhibit it as the spectrum of Ramond vacua:
\begin{equation}
\begin{array}{ll}
\dsp  L_0 \ket{\om,l ;(0)} = \frac{k}{4} \ket{\om,l ;(0)} & \dsp
 K^3_0 \ket{\om,l ;(0)} = \left(\frac{q(\om)}{2}+\frac{l}{2}-\frac{k}{2}\right)
     \ket{\om,l ;(0)} , \\
\dsp  \bar{L}_0 \ket{\om,l ;(0)} = \frac{k}{4} \ket{\om,l ;(0)} & \dsp 
 \bar{K}^3_0 \ket{\om,l ;(0)} = \left(\frac{\bar{q}(\om)}{2}
      +\frac{l}{2}-\frac{k}{2}\right)
     \ket{\om,l;(0)} .
\end{array}
\label{spectrum 4}
\end{equation}

   The analyses for the twisted sectors are more difficult.
 In general they are classified by Young tableaus
 $(n_1,\ldots,n_s)$ composed of $p$ boxes
 ($n_1 \geq \ldots \geq n_s >0$, $\dsp \sum_{i=1}^s n_i =p$). The sector 
 labeled by a tableau $(n_1,\ldots,n_s)$ can be decomposed into a set of
 $\bz_{n_i}$-twisted sectors. In other words it can be regarded as
 composed of $s$ glued strings of length $(n_1,\ldots,n_s)$. If we are
 interested in single particle states, we only have to consider the
 $\bz_m$-twisted sector, or the sector of a single glued string of
 length $m$.

   The superconformal algebra (SCA) suitably acting on the Hilbert space
 of the $\bz_m$-twisted sector is constructed in the following manner
 (see for example \cite{cyclic}): 
 First we make up the SCA 
 $\left\{L'_n,~ G^{\prime \alpha a}_n,~ K_n^{\prime \alpha\beta}\right\}$
 of the glued string variables as in the case of the single long string
 ${\cal M}_{\rm long}$ with $c=6k$. Second, we must mod out by
 $\bz_m$-action, and get the desired SCA $\hat{\cA}$ with $c=6mk$. 
 More explicitly, we should define the superconformal generators (in
 the Ramond sector) for the $\bz_m$-twisted sector as follows\footnote{
       The situation becomes more complicated for the SCA of NS sector,
       since we must define them separately according to whether $m$ is
       even or odd. 
       This is the main reason why here we work in the Ramond sector and
       make use of the spectral flow instead of dealing with the NS
       sector directly.}
\begin{equation}
\begin{array}{lll}
 \hat{L}_n & \df & \dsp \frac{1}{m} L'_{mn} 
 + \frac{k}{4}\left(m-\frac{1}{m}\right)\delta_{n0} \\
\hat{K}_n^{\al\beta} & \df & K_{mn}^{\al\beta\,'} \\
\hat{G}^{\al a}_n & \df & \dsp \frac{1}{\sqrt{m}} G^{\al a \, '}_{mn} .
\end{array}
\label{SCA twisted}
\end{equation}
In this expression for $\hat{L}_n$  the second term of the RHS corresponds
to the contribution from the Schwarzian derivative of the covering map
$w=z^m$.  

   We can now present the complete spectrum of chiral primaries for the
 $\bz_m$-twisted sector. The Ramond vacua in this sector should have the 
 same spectrum of weight and $R$-charge as those in the CFT of a single
 long string:
\begin{equation}
\begin{array}{ll}
\dsp  L'_0 \ket{\om,l ;(m)} = \frac{k}{4} \ket{\om,l ;(m)} & \dsp
 K^{3\, '}_0 \ket{\om,l ;(m)} 
= \left(\frac{q(\om)}{2}+\frac{l}{2}-\frac{k}{2}\right)
     \ket{\om,l ;(m)} , \\
\dsp  \bar{L}'_0 \ket{\om,l ;(m)} = \frac{k}{4} \ket{\om,l ;(m)} & \dsp 
 \bar{K}^{3 \, '}_0 \ket{\om,l ;(m)} = \left(\frac{\bar{q}(\om)}{2}
      +\frac{l}{2}-\frac{k}{2}\right)
     \ket{\om,l;(m)} .
\end{array}
\label{spectrum 5}
\end{equation}
 Hence, by the definitions\eqn{SCA twisted} we obtain the following
 relations:
\begin{equation}
\begin{array}{l}
\dsp  \hat{L}_0 \ket{\om,l ;(m)} = 
 \left\{\frac{k}{4m} + \left( \frac{km}{4}- \frac{k}{4m} \right) 
 \right\}\ket{\om,l ;(m)} = \frac{km}{4}\ket{\om,l ;(m)} \\  
\dsp  \hat{K}^{3}_0 \ket{\om,l ;(m)}  
= \left(\frac{q(\om)}{2}+\frac{l}{2}-\frac{k}{2}\right)
     \ket{\om,l ;(m)} , \\
\dsp  \bar{\hat{L}}_0 \ket{\om,l ;(m)} = \frac{km}{4} \ket{\om,l ;(m)} ,~~~~~ 
 \bar{\hat{K}}^{3}_0 \ket{\om,l ;(m)} = \left(\frac{\bar{q}(\om)}{2}
      +\frac{l}{2}-\frac{k}{2}\right)
     \ket{\om,l;(m)} .
\end{array}
\label{spectrum 6}
\end{equation}
 Finally we translate the above relations in the Ramond sector to those
 in NS sector by means of the spectral flow. As a result, the chiral
 primaries in ${\bz_m}$-twisted sector $\cO(\om,l;(m))$ have the
 following quantum numbers:
\begin{equation}
\begin{array}{c}
  \dsp
  h=j= \frac{q(\om)+l+k(m-1)}{2}, ~~~ 
  \bar{h}=\bar{j}= \frac{\bar{q}(\om)+l+k(m-1)}{2} \\
  (l=0,\cdots,k-2,~~~ m=1,\cdots, p)
\end{array}
\label{spectrum 7}
\end{equation}
 In this way we can reproduce almost all the chiral primaries of single
 particle type that are expected from the correspondence with the SCFT
 on $Sym^{pk}({ M}^4)$,{\em except for the absence of the following
 sequence of states}:
\begin{equation}
  h=j= \frac{q(\om)+km-1}{2}, ~~~ 
  \bar{h}=\bar{j}= \frac{\bar{q}(\om)+km-1}{2} ~~~(m=1,\cdots, p) .
\label{missing}
\end{equation}
 Notice that the case $m=1$ in \eqn{missing} is no other than the
 ``first missing state'' discussed in \cite{9903224}. There it
 was discussed that the absence of this state is related to the
 small-instanton singularity of the moduli space of $D1/D5$-brane bound
 states. We find its cousins in the ${\bz_m}$-twisted sectors
 $m >1$. Although we still have the missing states in the framework
 of Matrix string theory, there are ``not too many'' of them. Especially,
 the bound for the R-charge $\dsp \sim \frac{pk}{2}$ expected from
 the relation to $Sym^{pk}(M^4)$ sigma model is correctly reproduced. 

   In \cite{9903224} it was also pointed out that, in the single long
 string theory, the continuous spectrum appears above the threshold
 conformal weight, $\dsp \Delta_0 = \frac{(k-1)^2}{4k} \sim \frac{k}{4}$,
 which would make the counting of the chiral primaries above the
 threshold difficult. This result is based on the study of Liouville
 theory\cite{Seiberg-L}, in which it is claimed that we have two
 different types of states in quantum Liouville theory: one of them
 is a non-normalizable state which form a discrete spectrum, and the
 other is a normalizable state which form a continuous spectrum. The
 latter appears above the threshold, $\dsp \Delta_0\equiv \frac{Q^2}{8}$,
 where $Q$ is the background charge of Liouville theory. Here we shall
 show that the threshold value also becomes $p$-times larger when we
 consider the CFT of $p$ long strings, $Sym^p({\cal M}_{\msc{long}})$.

   It is easy to see this in the untwisted sector. We only have to find
 the normalizable state with the smallest conformal weight in this
 sector. Obviously this state has the form
 $\dsp \ket{\Delta_0}^{\otimes p}$, where $\ket{\Delta_0}$ is the
 normalizable state with the lowest energy in the single long string theory:
\begin{equation}
  L_0 \ket{\Delta_0} \sim \frac{k}{4}\ket{\Delta_0}.
\end{equation}   
 Therefore the threshold for the untwisted sector is equal to
 $\dsp \hat{\Delta}_0 \sim \frac{pk}{4}$.

   In order to calculate the threshold value for general twisted sectors
 we have to recall the definition of the Virasoro operator in the
 $\bz_m$-twisted sector\eqn{SCA twisted}. Then the threshold for the
 $\bz_m$-twisted sector becomes:
\begin{equation}
  \hat{\Delta}_0 \sim \frac{k}{4m} + \frac{k}{4}\left(m-\frac{1}{m}\right)
  = \frac{mk}{4} .
\end{equation}
 From these results we can conclude that the threshold value is the same 
 for all the twisted sectors characterized by arbitrary Young tableaus
 $(n_1,\ldots, n_s) \in Y_p$:
\begin{equation}
\hat{\Delta}_0 = \sum_{i=1}^s \, \frac{n_ik}{4} = \frac{pk}{4}.
\end{equation}

   We make one comment. It is known that the spectrum of chiral
 primaries in $Sym^{pk}(M^4)$ sigma model agrees with that of
 supergravity below the bound $\dsp \sim \frac{pk}{4}$. This claim was
 first proved in the case of $M^4= K3$ by analyzing the elliptic
 genus\cite{de Boer}, and more recently it has been proved in the case
 of $M^4=T^4$ by analyzing the ``new SUSY index''\cite{MMS}. 
 The value of the upper bound is the same, $\dsp \frac{pk}{4}$.
 Although apparently the origins of the bound are different between the
 analyses of \cite{de Boer,MMS} and ours, there might be some
 relationship between the two. This is because supergravity on an
 $AdS_3 \times S^3 \times M^4$
 background has only the discrete spectrum, so that the emergence of
 continuous spectrum may lead to a failure in the correspondence between
 the description of supergravity and the boundary SCFT.

~

\section{Conclusions and Discussions}
\cleqn

   In this paper we studied the multi-string system on 
 $AdS_3\times S^3\times M^4$ in the framework of Matrix string theory.
 Among other things, we have successfully presented a unified framework
 of the various short and long string sectors. Although gauge field in
 two dimensions has no local physical degrees of freedom, it plays an
 extremely important role, namely, the VEV of electric field on the
 worldsheet distinguishes between the short and long string sectors. 

   Now we would like to mention some questions which could be discussed
 in future works.

   In section 3 we have shown the T-duality between the Matrix string
 action and the $AdS_3\times S^3$ string action, as far as their
 bosonic parts are concerned. It would be interesting to investigate
 further and see the full T-duality. Of course, some fermions are
 missing on the gauge-theory side (which would correspond to the
 superpartner of bosonic $SL(2,R)$ currents on the superstring theory.)
 Perhaps this aspect will be better understood when we come to a deeper
 understanding of the relation between the $U(1)$ gauge symmetry of
 Matrix string theory and the reparametrization invariance of the
 superstring theory. To this aim the formulation of Matrix theory in a
 manifestly covariant manner may be useful. Some approaches to this subject
 were given in \cite{FO}.

   In section 5 we discussed the problem of the missing chiral primaries
 by the analysis of the CFT of $p$ long strings. Although we still have
 some states missing, as was pointed out in \cite{9903224}, the
 correspondence with the $Sym^{pk}(M^4)$ sigma model is not ``too bad''.
 In particular, by taking account of the various glued strings we have
 reproduced the bound $\dsp \sim \frac{pk}{2}$
 which is precisely the same as is expected from the analysis of the
 $Sym^{pk}(M^4)$ sigma model.   

   However, there seems to be a further subtlety regarding the
 spectrum of the multi-particle chiral primaries, for our construction
 seems to yield the spectrum interpretable as the cohomology of
 $Sym^p(Sym^k(M^4))$, not of $Sym^{pk}(M^4)$. In other words there
 should be an U-duality transformation which changes the value of $k$
 and $p$, while preserving their product. At present it is not clear
 whether our construction in the framework of Matrix string theory has
 this duality symmetry. In any case, we will need further study to get
 a complete understanding of this issue.

   The problem of the threshold for the continuous spectrum is also
 interesting. We have shown that our analysis of the CFT of $p$ long
 strings yields a threshold $p$-times larger, $\dsp \sim \frac{kp}{4}$.
 This is the same as the threshold obtained in \cite{de Boer,MMS}. It was 
 discussed there that the correspondence between the supergravity in
 bulk and the boundary SCFT fails above this threshold. Perhaps it may
 be interesting to study the relation between this issue and the physics
 of three-dimensional black holes, as the above threshold coincides
 with the energy at which the first excited BTZ black hole (the massless
 BTZ black-hole) appears.

~

\medskip
\noindent {\Large\bf Acknowledgment}
\medskip

   We would like to thank T. Eguchi for discussions and useful comments.
 We would also like to thank B.D. Bates for careful reading of the
 manuscript. K.H is supported in part by JSPS Research Fellowships,
 and also Y.S is supported in part by the Grant-in-Aid  from
 the Ministry of Education, Science and Culture, Priority Area:
 ``Supersymmetry and Unified Theory of Elementary Particles"
 $(\sharp 707)$.

\end{document}